\documentclass[fleqn,usenatbib]{mnras}

\usepackage[T1]{fontenc}
\usepackage{ae,aecompl}
\usepackage[dvipsnames]{xcolor}

\usepackage{graphicx}	
\usepackage{amsmath}	
\usepackage{amssymb}	
\usepackage{ctable} 
\usepackage{enumitem}
\usepackage{bm}

\newcommand{\hi}{H\textsc{i}\ }
\newcommand{\hinospace}{\textrm{H\textsc{i}}}
\newcommand{\de}{\text{d}}
\newcommand{\fnl}{f_{\rm NL}}
\newcommand{\Tbar}{\overline{T}_\hinospace}
\newcommand{\kFG}{k_\parallel^\text{FG}}

\renewcommand{\P}{\mathcal{P}}
\newcommand{\T}{\mathcal{T}}
\newcommand{\W}{\mathcal{W}}

\newcommand{\secref}[1]{\hyperref[#1]{Section~\ref*{#1}}}
\newcommand{\appref}[1]{\hyperref[#1]{Appendix~\ref*{#1}}}

\title[Investigating PNG and 21cm foreground degeneracies]{The degeneracy between primordial non-Gaussianity and foregrounds in 21cm intensity mapping experiments}

\author[S. Cunnington et al.]{
Steven Cunnington,$^{1}$\thanks{E-mail: s.cunnington@qmul.ac.uk}
Stefano Camera,$^{2,3,4}$
Alkistis Pourtsidou$^{1,4}$
\\
$^{1}$School of Physics and Astronomy, Queen Mary University of London, Mile End Road, London E1 4NS, UK\\
$^{2}$Dipartimento di Fisica, Universit\`{a} degli Studi di Torino, via P. Giuria 1, 10125 Torino, Italy\\
$^{3}$
INFN – Istituto Nazionale di Fisica Nucleare, Sezione di Torino, via P. Giuria 1, 10125 Torino, Italy\\
$^{4}$Department of Physics \& Astronomy, University of the Western Cape, Cape Town 7535, South Africa\\
}

\date{Accepted XXX. Received YYY; in original form ZZZ}

\pubyear{2020}

\begin{document}
\label{firstpage}
\pagerange{\pageref{firstpage}--\pageref{lastpage}}
\maketitle

\begin{abstract}
Potential evidence for primordial non-Gaussianity (PNG) is expected to lie in the largest scales mapped by cosmological surveys.  Forthcoming 21cm intensity mapping experiments will aim to probe these scales by surveying neutral hydrogen (\hinospace) within galaxies. However, foreground signals dominate the 21cm emission, meaning foreground cleaning is required to recover the cosmological signal. The effect this has is to damp the \hi power spectrum on the largest scales, especially along the line-of-sight. Whilst there is agreement that this contamination is potentially problematic for probing PNG, it is yet to be fully explored and quantified. In this work we carry out the first forecasts on $\fnl$ that incorporate simulated foreground maps that are removed using techniques employed in real data. Using an MCMC analysis on an SKA1-MID-like survey, we demonstrate that foreground cleaned data recovers biased values ($\fnl = -102.1_{-7.96}^{+8.39}$ [68\% CL]) on our $\fnl=0$ fiducial input. Introducing a model with fixed parameters for the foreground contamination allows us to recover unbiased results ($\fnl = -2.94_{-11.9}^{+11.4}$). However, it is not clear that we will have sufficient understanding of foreground contamination to allow for such rigid models. Treating the main parameter $\kFG$ in our foreground model as a nuisance parameter and marginalizing over it, still recovers unbiased results but at the expense of larger errors ($\fnl = 0.75^{+40.2}_{-44.5}$), that can only be reduced by imposing the Planck 2018 prior. Our results show that significant progress on understanding and controlling foreground removal effects is necessary for studying PNG with \hi intensity mapping. 
\end{abstract}

\begin{keywords}
cosmology: large-scale structure of Universe -- cosmology: theory -- cosmology: observations -- radio lines: general
\end{keywords}

\section{Introduction}

The concordance cosmological model, $\Lambda$CDM, provides a very successful framework to explain our cosmological observations. However, several open questions remain. Along with the requirement of dark energy and dark matter, a mechanism for a period of rapid accelerated expansion immediately after the Big Bang is also required. To provide this, models of \emph{cosmological inflation} have been developed \citep{PhysRevD.23.347,LINDE1982389}, which successfully explain many problems with the standard hot Big Bang model of cosmology. Furthermore, quantum vacuum fluctuations sourced by inflation provide the perturbations required to produce the cosmic microwave background (CMB) anisotropies and large-scale structure of the Universe \citep{Starobinsky:1982ee, Mukhanov:1982nu}.

Along with the necessity of the inflationary period to explain observations, comes a plethora of suggested inflationary models \citep{Bartolo:2004if}. The simplest model of inflation postulates a single self-interacting scalar field that drives the accelerated expansion \citep{Seery:2005wm}. A fundamental prediction of single-field inflation is that the primordial fluctuations in our Universe are Gaussian to a good approximation \citep{Maldacena:2002vr}.

In this work we focus on constraining the primordial non-gaussianity (PNG) parameter, $f_\mathrm{NL}$, which quantifies the departure from Gaussianity \citep{Komatsu:2001rj}. In particular, we focus on so-called local-type PNG. This probe is sensitive to multi-field models of inflation and as such, an experimental confirmation of $\fnl \neq 0$ would rule-out single-field inflation in favour of a multi-field scenario \citep{Creminelli:2004yq}.

Currently, the best constraint on $\fnl$ comes from measuring the bispectrum of the CMB, which has achieved a constraint of $\fnl = 0.9 \pm 5.1$ \citep{Akrami:2019izv}. However, the information available from this approach is close to saturation and instead probing $\fnl$ with large-scale structure probes is seen as the most likely method of achieving $\sigma(\fnl)\sim 1$ \citep{Sefusatti:2012ye, Alvarez:2014vva,Raccanelli:2014awa}. Investigations looking to constrain $\fnl$ using large-scale structure surveys have been undertaken \citep{Slosar:2008hx, Ross:2012sx,Leistedt:2014zqa,Castorina:2019wmr}, however, it is from next-generation surveys e.g.\ LSST\footnote{\href{https://www.lsst.org/}{lsst.org}}, Euclid\footnote{\href{https://www.euclid-ec.org/}{euclid-ec.org}}, DESI\footnote{\href{https://www.desi.lbl.gov/}{desi.lbl.gov}} \citep{Abell:2009aa,laureijs2011euclid,Levi:2013gra} where $\fnl$ constraints should become competitive \citep{Byun:2014cea,Ballardini:2016hpi}.

In addition to the above surveys, the Square Kilometre Array (SKA)\footnote{\href{https://www.skatelescope.org/}{skatelescope.org}} \citep{Bacon:2018dui} offers an alternative strategy since it maps large-scale structure using radio telescopes that will also be able to probe PNG \citep{Camera:2013kpa,Camera:2014bwa,Alonso:2015uua,Gomes:2019ejy}. This is most efficiently done using neutral hydrogen (\hinospace) intensity mapping which maps the combined, unresolved 21cm emission from galaxies \citep{Bharadwaj:2000av,Battye:2004re,Chang:2007xk}. By using the single-dish approach \citep{Battye:2012tg, Bull:2014rha}, each dish in the array can scan the sky independently providing a rapid strategy for mapping large volumes with good signal-to-noise. 

A major issue with \hi intensity mapping observations is contamination from diffuse foregrounds: 21cm foregrounds are caused by astrophysical processes emitting radiation in the same frequency range as the \hi signal. These can be orders of magnitudes larger than the \hi cosmological signal, but they can be removed using techniques similar to the ones used in CMB experiments \citep{Chapman:2012yj, Wolz:2013wna,Shaw:2014khi,Alonso:2014dhk,Anderson:2017ert,Carucci:2020enz}. This unfortunately causes a contamination, mainly to large cosmological modes parallel to the line-of-sight (LoS), leading to anisotropic signal loss. Understanding these effects on cosmological analyses is paramount \citep{Cunnington:2019lvb,Cunnington:2020mnn,Shi:2020aaj}. In addition, for radio telescopes the transverse (angular) resolution is effectively determined by the baseline of the receivers, i.e.\ the maximum separation of incident radiation on the receivers. By opting for a single-dish approach, this baseline is limited to the diameter of the dish. Thus, the smaller the diameter of the dish, the larger the beam and the poorer the resolution. The beam can therefore have direct effects on small-scale transverse modes, and understanding the contamination from this, is also important \citep{Cunnington:2020mnn}.

Whilst opting for a single-dish approach with a wide beam does degrade effective resolution and renders certain transverse modes inaccessible \citep{Villaescusa-Navarro:2016kbz}, it is the largest scales which are most sensitive to $\fnl$ and therefore the beam is not expected to have much impact. Further benefits provide cause for optimism towards using \hi intensity mapping to probe PNG. It has been shown that poorly understood astrophysical processes connected with \hi should not have a significant impact on the measurement of cosmological parameters with intensity mapping \citep{Padmanabhan:2018llf} and they should not bias the measurement of $\fnl$ \citep{Camera:2019iwy}. Also, relativistic effects in lensing magnification which ordinarily bias PNG probes \citep{Camera:2014sba,Wang:2020ibf} are not present in intensity mapping \citep{Hall:2012wd}.

However, the effect on $\fnl$ measurements from foreground contamination which erodes information on large scales, is yet to be fully explored despite agreement that it could be quite problematic \citep{Camera:2013kpa,Xu:2014bya,Alonso:2015uua, MoradinezhadDizgah:2018lac}. There has been previous work investigating systematic effects on probing PNG. For example, there has been studies into the effect of contamination from line interlopers in spectroscopic surveys \citep{Lidz:2016lub,Cheng:2016yvu} many of which in the context of PNG \citep{Pullen:2015yba, Addison:2018xmc,Gebhardt:2018zuj, Gong:2020lim}. \citet{Kalus:2018qsy} investigated removing contaminated modes for galaxy survey data and the impact this has on probing PNG. There have also been attempts to approximately incorporate 21cm foreground contamination into Fisher matrix forecasts \citep{Lidz:2013tra,Karagiannis:2019jjx}.

In this work, for the first time, we assess the ability of \hi intensity mapping experiments to probe $\fnl$ by producing simulated data sets inclusive of foreground contamination and employing foreground removal algorithms together with modelling and parameter estimation techniques emulating a real data analysis. This is the most robust way to quantify the potential effects on the determination of $\fnl$ from foreground removal systematics. 

For the purposes of our forecasts, we construct simulations that aim to emulate a \hi cosmological signal detected by an SKA1-MID Band 1 survey \citep{Bacon:2018dui}. We add simulated foregrounds into this signal and clean these with a foreground removal algorithm to investigate the impact from this reconstruction process. By constructing a model of this contamination with free parameters which can be marginalized over, we attempt to recover the fiducial (true) $\fnl$ value from our simulations and the associated measurement errors. This serves as a more realistic forecast of future PNG constraints using \hi intensity mapping in the presence of foregrounds.

The paper is outlined as follows. In \secref{PNGSec} we formalise the study of probing PNG in the context of \hi intensity mapping. In \secref{Methodology} we outline our methodology including details on our simulated data. \secref{MCMCresults} provides the results we obtained from the Monte Carlo Markov Chains (MCMC) analysis of the data under different scenarios. We further discuss our results in \secref{Discussion} and conclude in \secref{Conclusion}.

\section{PNG with Intensity Mapping}\label{PNGSec}

In the absence of PNG, the large scale \hi power spectrum, as a function of redshift $z$, wave vector $\bm k$, and the cosine of the angle $\theta$ between $\bm k$ and the LoS i.e.\ $\mu\equiv \cos(\theta)$, is given as
\begin{equation}
\label{eq:PKnofNL}
    P_\hinospace(k,\mu,z) = \Tbar(z) ^2 \left[b_\hinospace(z) + f(z)\mu^{2}\right]^{2} P_\mathrm{m}(k,z) \, .
\end{equation}
Here, $\Tbar$ is the mean \hi temperature of the field (proportional to the \hi density $\Omega_\hinospace$), $b_\hinospace$ is the linear bias, and $P_\mathrm{m}$ is the matter power spectrum. The $f\mu^2$ term accounts for linear redshift space distortions (RSD) where $f$ is the linear growth rate of structure, implemented to model the anisotropies caused from RSD \citep{Kaiser:1987qv}. 

The presence of PNG of most types leads to a strong scale-dependent correction to the linear bias \citep{Dalal:2007cu}. In this way, large-scale structure surveys can be used to probe $\fnl$. For the local-type PNG we focus on, the $\fnl$-dependent correction term scales as $k^{-2}$, it is thus at sufficiently large scales (small-$k$) where signs of PNG should manifest. It is also reasonable to assume that these large scales should remain uncontaminated by the nonlinear growth of collapsed structures.

A modification to the scale-independent Gaussian bias on large scales can be written as
\begin{equation}
    b_\hinospace(z) \rightarrow b_\hinospace(z)+\Delta b_\hinospace(z, k)\,,
\end{equation}
where the scale-dependent, non-Gaussian correction is given by \citep{Dalal:2007cu, Slosar:2008hx}
\begin{align}\label{eq:ScaleBias}
     \Delta b_\hinospace(z, k) &=  \left[b_\hinospace(z)-1\right]\frac{3\,\Omega_\mathrm{m}\,H_0^2\,\delta_\mathrm{c}}{c^2\,k^2\,T(k)\,D(z)}\,\fnl\, ,\\
     \label{eq:ScaleBias2}&\equiv \widetilde{\Delta b}_\hinospace(z, k)\,\fnl,
\end{align}
where $\delta_\mathrm{c} \simeq 1.686$ is the critical matter density contrast for spherical collapse, $T(k)$ is the matter transfer function adopting the convention $T(k\rightarrow 0)=1$, and lastly in \autoref{eq:ScaleBias}, we have the growth function defined as
\begin{equation}
    D(z)=\frac{5}{2} \Omega_\mathrm{m}\,H_0^2\,H(z)\int_{z}^{\infty}\frac{1+z^{\prime}}{H^3(z^{\prime})}\,\de z^{\prime}\, ,
    \label{eq:growth_factor}
\end{equation}
which is normalised to unity at $z=0$. In the second line of \autoref{eq:ScaleBias2}, we have explicitly factorised $\fnl$ out, for reasons that will become clear later on. Therefore the \hi power spectrum, including the possibility for non-zero $\fnl$, is given as
\begin{multline}\label{eq:HIPkwfNL}
    P_\hinospace(k,\mu,z) =\\ \Tbar ^2 \left[b_\hinospace(z) + \widetilde{\Delta b}_\hinospace(z,k)\fnl +  f(z)\mu^{2}\right]^{2} P_\mathrm{m}(k,z) \, .
\end{multline}

\subsection{Impact of 21cm Foregrounds on PNG}\label{FGeffectonPNG}

\begin{figure}
	\centering
  	\includegraphics[width=\columnwidth]{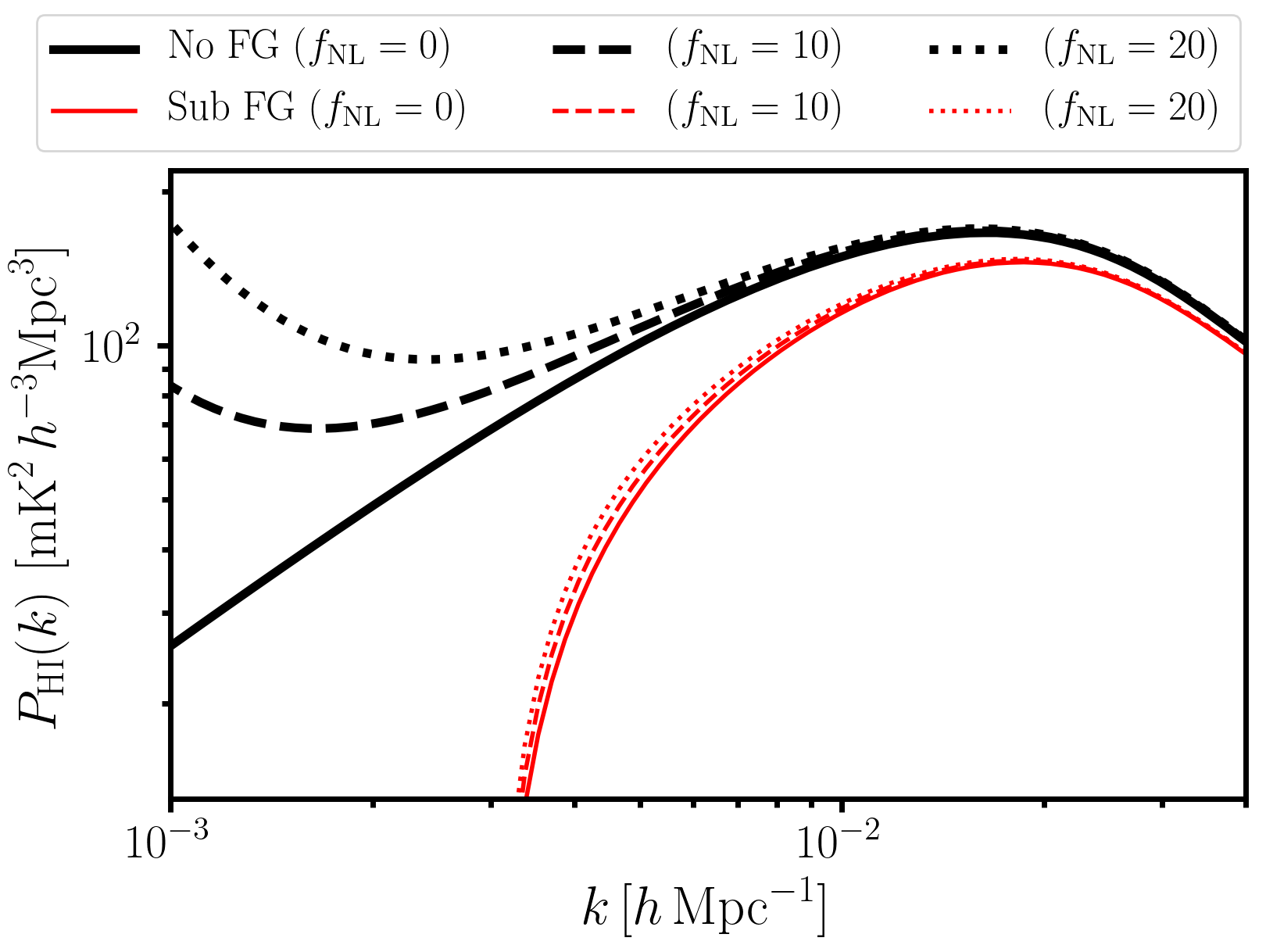}
    \caption{Model power spectra with different amounts of $\fnl$. \textit{Black-thick} lines show the foreground-free case, whereas the \textit{red-thin} lines shows the result from excluding all large parallel modes with $k_\parallel<3\times 10^{-3}\,h\,\text{Mpc}^{-1}$ as a simple model of foreground contamination. In this latter case, the impact from $\fnl$ is far more suppressed.}
\label{fig:ModelfNL}
\end{figure}

Probing \hi in the late, post-reionization Universe requires making observations in frequency ranges $350\,\text{MHz}<\nu<1420\,\text{MHz}$ (approximately redshifts of $z<3$). Conducting these observations with the intensity mapping method means data will also contain diffuse emission from other astrophysical processes within the same frequency range. This unwanted, additional emission is referred to as a \textit{foreground}. Due to the inherently weak cosmological \hi signal ($\Tbar \simeq 0.079\,\text{mK}$ at $700\,\text{MHz}$), the foregrounds dominate observations, for example the Galactic synchrotron signals can be several orders of magnitude greater ($\overline{T}_\mathrm{gsync} \simeq 17,000\,\text{mK}$). It is therefore imperative that astrophysical foregrounds are removed with minimal residuals to allow \hi intensity mapping to be a competitive and unbiased cosmological probe.

Typically this is done by utilising the fact that the majority of the signal from foregrounds will be a smooth continuum through frequency, whereas the \hi cosmological signal is expected to be stochastic. This can be used as a distinguishing feature to separate foregrounds and then remove them. There is detailed literature on foreground removal in the context of intensity mapping \citep{DiMatteo:2001gg,Oh:2003jy,Wang:2005zj,Morales:2005qk,Liu:2011ih, Wolz:2013wna,Shaw:2014khi,Alonso:2014dhk,Cunnington:2019lvb}, which we refer the reader to for a more comprehensive discussion.

Assuming the matter transfer function converges as $T(k)\rightarrow 1$ for small-$k$, it is then apparent from the $1/k^2$ factor in \autoref{eq:ScaleBias} that evidence for a scale-dependent bias, and hence for PNG, is strongest on the largest scales (small-$k$). A challenge therefore arises when trying to constrain $\fnl$ using \hi intensity mapping. Foreground cleaning will inevitably remove large-scale modes from the \hi intensity map, potentially destroying the information required to detect PNG signals in large-scale structure. This is because we expect foregrounds to have relatively smooth fluctuations along the LoS, and a blind foreground clean will primarily remove cosmological modes at small $k_\parallel$ since these are the ones that will be indistinguishable from the foregrounds. 

A simple approach to modelling the damping caused by a foreground clean is to apply a $k_\parallel$ cut on the power spectrum $P_\hinospace(k)$ to emulate lost large-scale modes along the LoS. As an early demonstration of the potential difficultly of probing $\fnl$ with this assumption of lost modes, we show various power spectra in \autoref{fig:ModelfNL} with differing $\fnl$. The thick black lines are without any loss of modes and the thinner red lines are where all modes with $k_\parallel < 3\times 10^{-3}\,h\,\text{Mpc}^{-1}$ are lost, to emulate the foreground contamination. It is immediately evident from \autoref{fig:ModelfNL} that by losing these modes the sensitivity to $\fnl$ is greatly reduced. Therefore, if this assumption is correct and all $k_\parallel$ modes below a certain limit are lost in a foreground clean, then measuring $\fnl$ with intensity mapping could be severely problematic. Indeed, as we later demonstrate in more detail, we find that $\fnl$ measurements are hugely biased if the effects from foreground cleaning are not sufficiently modelled.

\subsection{Modelling Intensity Mapping Systematics}\label{ModellingIMsystematics}

Including the possibility of a non-zero $\fnl$, the \hi intensity mapping power spectrum is given by (where we have omitted redshift dependence for simplicity)
\begin{multline}\label{eq:HIPkwithFGbeamModel}
    P_\hinospace(k, \mu) = P_\mathrm{N}\,\widetilde{B}_\mathrm{b}^2(k,\mu)\,\widetilde{B}_\mathrm{FG}(k,\mu) \quad + \\ \overline {T}_\hinospace^2 \left(b_\hinospace + \widetilde{\Delta b}_\hinospace\,\fnl +f\mu^{2}\right)^{2} P_\mathrm{m}(k)\,\widetilde{B}_\mathrm{b}^2(k,\mu)\,\widetilde{B}_\mathrm{FG}(k,\mu)\, ,
\end{multline}
which is an extension of \autoref{eq:HIPkwfNL} to include an instrumental noise model $P_\mathrm{N}$ (discussed later in \secref{ThermalNoise}) and damping models for both foreground removal ($\widetilde{B}_\mathrm{FG}$) and the radio telescope beam ($\widetilde{B}_\mathrm{b}$), which smooths large transverse modes, $\bm k_\perp$. The tildes above these quantities are used to denote that they are both functions in Fourier space. Whilst the effect from the beam is still large enough to warrant consideration, in this particular study which relies on information on the largest scales, it is less likely to have a significant impact (see also \citet{Camera:2019iwy}). Furthermore, the beam for each particular instrument is likely to be more understood and thus easier to model than the effects caused by foregrounds. For the beam, since we know our data has been smoothed with a symmetric Gaussian (outlined in \secref{InstEffectsSec} where we discuss our simulations), we can be confident in modelling these effects correctly with
\begin{equation}
    \widetilde{B}_{\mathrm{b}}(k, \mu)=\exp \left[-\frac12k^{2} R_\mathrm{b}^{2}\left(1-\mu^{2}\right)\right]\,,
\end{equation}
where $R_\mathrm{b} = \chi(z) \theta_{\mathrm{FWHM}} /(2 \sqrt{2 \ln 2})$, $\theta_\mathrm{FWHM}$ is the full-width-half-maximum of the beam in radians, and $\chi(z)$ the comoving distance to redshift $z$. In studies where the beam is thought to have a more relevant effect e.g.\ baryon acoustic oscillations studies, the beam size $R_\mathrm{b}$ can be treated as a free parameter in the model to reflect possible uncertainty in its behaviour \citep{Villaescusa-Navarro:2016kbz}. 

For modelling the foregrounds, we need $\widetilde{B}_\mathrm{FG}$ to be some function that damps large modes. In \citet{Cunnington:2020mnn}, a free parameter $k_\parallel^\text{FG}$ was introduced representing some parallel wave-vector scale and it was shown that a cut below some tuned value for $k_\parallel^\text{FG}$ can be an effective way of modelling the foreground removal effects on simulated data in the context of multipole expansion of the power spectrum. The foreground removal effects are assumed unrecoverable for modes below $k_\parallel^\text{FG}$ yet if no foregrounds are present, or a perfect clean is carried out, then we have $k_\parallel^\text{FG} = 0$. 
However, in this work where most of the constraining power comes from the large scales, we found this approach alone is not optimal. Whilst the very largest modes (smallest $k_\parallel$) are completely lost, the remaining modes are more gradually damped and we therefore use the below phenomenological function to model the foreground contamination:
\begin{equation}\label{eq:FGmodel}
    \widetilde{B}_\mathrm{FG}(k,\mu) = \alpha_\mathrm{FG}\, \Theta_\mathrm{FG}(k_\parallel)\,\left(1 - \exp\left[-\left(\frac{k_\parallel}{k^\text{FG}_\parallel}\right) \right]\right)\, .
\end{equation}
Here the bracketed exponential term, performs a damping to modes as a function of $k_\parallel$, where smaller $k_\parallel$ are more damped. $\Theta_\mathrm{FG}(k_\parallel)$ represents a Heaviside function which eliminates the smallest $k_\parallel$ modes accessible by the survey, essentially removing the smallest $k_\parallel$ bin since  these are assumed entirely foreground contaminated. This can be defined by
\begin{equation}\label{eq:Heaviside}
    \Theta_\mathrm{FG}(k_\parallel) =
    \begin{cases}
    0 & k_\parallel < k^\text{min}_\parallel \\
    1 & k_\parallel > k_\parallel^\text{min} \, .
    \end{cases}
\end{equation}
where $k^{\min}_\parallel=2\pi/L_\mathrm{z}$, and $L_\mathrm{z}$ is the depth of the survey in $\text{Mpc}/h$. Since $k^{\min}_\parallel$ is defined by the survey and binning strategy, the only free parameters in this model are $k^\text{FG}_\parallel$ and $\alpha_\mathrm{FG}$ where the latter is needed to globally damp the power spectrum independently of $k_\parallel$ or $\bm k_\perp$ and we find $\alpha_\mathrm{FG} = 0.97$ is sufficient for our data. Without $\alpha_\mathrm{FG}$ we obtained biased results for $\Tbar$ since our model was not accounting for slight damping from foregrounds on very small scales.

An alternative approach to modelling the damping caused by foreground contamination is to reverse the effects by constructing a foreground transfer function $T_\mathrm{FG}(k_\perp,k_\parallel)$, where $k_\perp \equiv |\bm k_\perp| = \sqrt{k_{\perp,1}^2 + k_{\perp,2}^2}$ and $k_{\perp,1}$, $k_{\perp,2}$ are the transverse modes. This is typically constructed in 2D ($k_\perp,k_\parallel$) to account for the anisotropic nature of the effects of a foreground clean \citep{Switzer:2015ria}. The transfer function is then applied to the measured power spectrum. Most simply the foreground transfer function can be given as
\begin{equation}
    T_\mathrm{FG}(k_\perp,k_\parallel) = \left\langle \frac{P_\mathrm{clean}(k_\perp,k_\parallel)}{P_{\hinospace,\text{noFG}}(k_\perp,k_\parallel)} \right\rangle\, ,
\label{eq:Transferfunc}
\end{equation}
where $P_{\hinospace,\text{noFG}}$ is for foreground-free intensity maps and $P_\mathrm{clean}$ is for maps with foregrounds added and then cleaned. The angled brackets denote an averaging over a number of realizations. In an analysis on real current data, which are dominated by noise and systematic effects, the procedure is more complicated \citep{Switzer:2015ria}. More specifically, the transfer function is determined by injecting the real data with simulated (mock) \hi signal data, performing a foreground clean, then determining how much signal is lost as a result. A proxy for this is e.g.\ taking the power spectrum of the cross-correlation of the cleaned combined (data $+$ mock) data with the mock data, and dividing by the mock power spectrum \citep[for details, see][]{Masui:2012zc,Switzer:2013ewa,Anderson:2017ert}. A transfer function in this context can also account for the telescope beam. However, in this work constructing a foreground transfer function using simulations and then applying it to data also produced with simulations is circular. Furthermore, by attempting to model the foregrounds, as we choose to, free-parameters can either be fitted for or marginalised over. Whereas using a transfer function, a large amount of faith has to be placed into the method of its construction and that the loss of signal in the simulations is an accurate representation of the loss of signal in the data. For our primary aim in this work, $\fnl$, and its degeneracies with foreground removal effects, this can be very important so we opt for the free-parameter approach. 

\begin{figure}
	\centering
  	\includegraphics[width=\columnwidth]{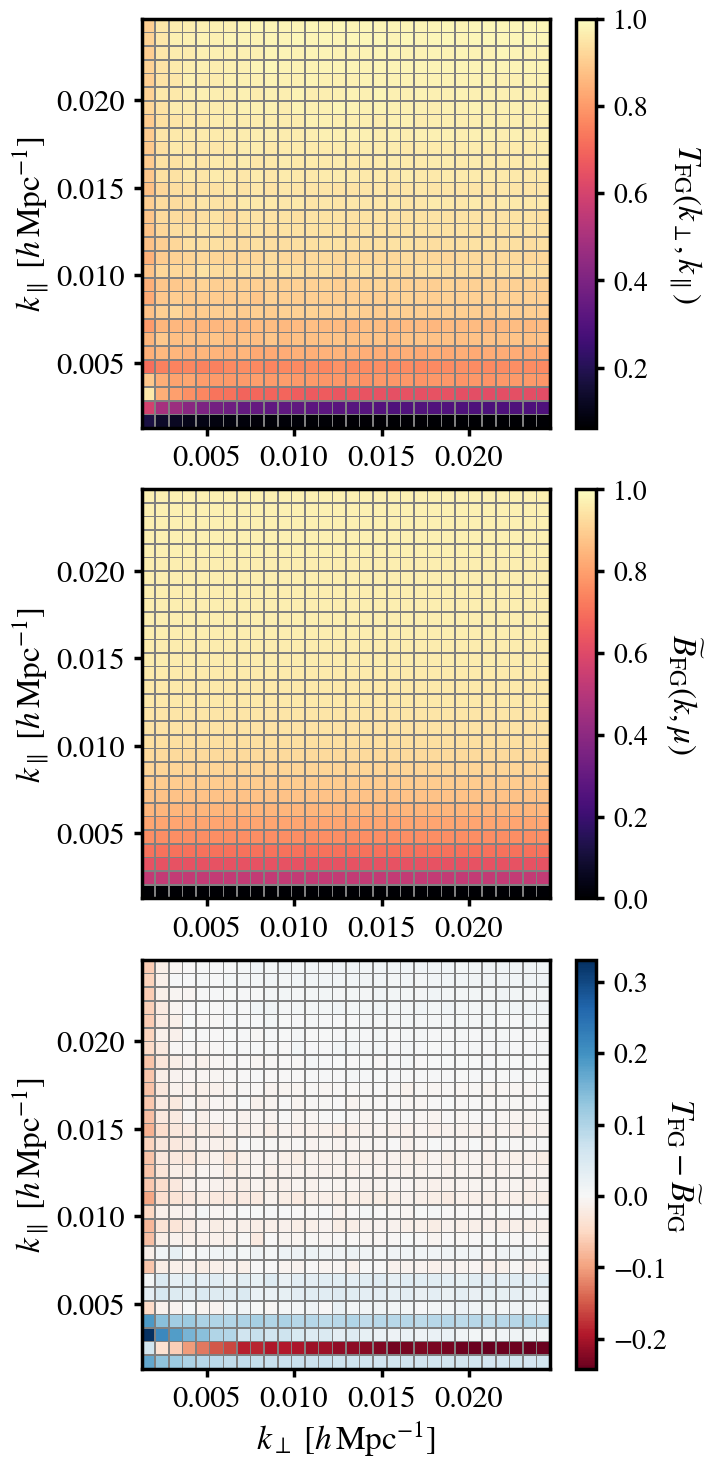}
    \caption{\textit{Top}-panel measured foreground transfer function (\autoref{eq:Transferfunc}) averaged over 100 \hi intensity map simulations. \textit{Middle}-panel shows the foreground contamination model (\autoref{eq:FGmodel}) using $k^\text{FG}_\parallel = 3.1\times 10^{-3}\,h\,\text{Mpc}^{-1}$. \textit{Bottom}-panel shows the residuals between the two with red (blue) regions representing where our model is under- (over-) estimating the damping from foreground contamination.}
\label{fig:TransfervsBfg}
\end{figure}

Since we know the true (foreground-free) power spectrum for our simulated data, it is possible to measure the foreground transfer function, which we can then compare to our generalised model to check it has the desired characteristics. \autoref{fig:TransfervsBfg} top-panel shows the measured foreground transfer function (calculated using \autoref{eq:Transferfunc}) for 100 simulated \hi intensity maps (simulation details are discussed in \secref{Methodology}). The middle panel shows our foreground model (\autoref{eq:FGmodel}) where we use $k^\text{FG}_\parallel = 3.1\times 10^{-3}\,h\,\text{Mpc}^{-1}$ for the free parameter value which we fit by eye in an attempt to match the measured data. The bottom row of $k_\parallel$ values in the measured transfer function motivates the need for the Heaviside part of the foreground model, as it appears the foreground removal is almost entirely eliminating the modes in this bin. Most of the remaining bins appear to be reasonably well matched by the exponential damping function as shown by the residuals in the bottom panel. There is slight evidence of some perpendicular dependence in the measured transfer function but modelling this would involve the introduction of at least one more free parameter which we choose to avoid.

The two top panels displayed in \autoref{fig:TransfervsBfg} should ideally match and the result is encouraging. Barring some disagreement on large scales, which could only be further improved by introducing more fitting parameters, \autoref{fig:TransfervsBfg} demonstrates that we can emulate the effects of foreground removal in our simulations with a 2-parameter model. With this at hand, we can proceed to experiment with constraining the value of $\fnl$ with \hi intensity mapping and investigate if a degeneracy exists between its measurement and contamination from a foreground clean.

\section{Methodology}\label{Methodology}

We construct our simulated data with the aim of emulating an SKA1-MID Band 1 intensity mapping survey. This instrument is expected to be able to deliver very competitive constraints on PNG due to the large scales it will probe with a wide and deep survey \citep{Bacon:2018dui}. We outline the survey parameters we have assumed in \autoref{SKASurveyTable}.

We chose to use the full redshift range of Band 1 for our data. Using such a wide redshift range is a reasonably common approach in experiments that aim to constrain $\fnl$,  since accessing the largest modes is hugely important \citep[e.g.][]{Mueller:2017pop}. Whilst the conventional approach of using smaller redshift bins allows for a more robust assumption of slowly evolving cosmology and other parameters, it restricts the maximum scale that can be probed, thus proving non-optimal for probing PNG. Furthermore, 21cm foreground contamination is mitigated for larger frequency (or redshift) ranges, since the increased depth allows the foreground clean to perform more efficiently and remove fewer cosmological modes \citep{Lidz:2013tra}. It is therefore important for this particular approach to maximise the redshift range of the data set.

It is possible to construct redshift weighting schemes that allow the analysis to be done on the full surveys redshift range \citep[e.g.][]{Zhu:2014ica,Mueller:2017pop}. \citet{Castorina:2019wmr} adopted this approach on eBOSS DR14 data by utilising optimal weights to account for redshift evolution in the survey which spanned the range $0.8 \leq
z \leq 2.2$. Using an analogous weighting scheme for the purpose of \hi intensity mapping surveys is certainly possible. For example, \citet{Blake:2019ddd} recently generalised the optimal weighting scheme by \citet{Feldman:1993ky} for \hi intensity mapping experiments and cross-correlations with optical galaxy surveys. However, since this is beyond the scope of this study (here we want to concentrate on the systematic bias coming from foreground removal), we assume a constant effective redshift of $z_\mathrm{eff}=1.675$ equal to the central redshift of the SKA Band 1. Therefore we assume all relevant parameters (e.g.\ the bias, average field temperature, beam size etc.) are redshift (or frequency) independent and evaluate them at this effective redshift. 
\begin{table}
	\centering
	\begin{tabular}{lcc} 
		\multicolumn{3}{c}{\textbf{SKA1-MID Survey Parameters}} \\
		\hline
        Band 1 Range & $(\nu_\mathrm{min},\nu_\mathrm{max})$ & $(350,1050)\,\text{MHz}$ \\
        Redshift Range & $(z_\mathrm{min},z_\mathrm{max})$ & $(0.35,3)$ \\
        Sky Fraction Coverage & $f_\mathrm{sky}$ & $0.52$ \\
        Channel Width & $\delta\nu$ & $1\,\text{MHz}$ \\
        Observation Time & $t_\mathrm{obs}$ & $10,000\,\text{hrs}$ \\
		Number of Dishes & $N_\mathrm{dish}$ & $197$ \\
        Dish Diameter & $D_\mathrm{dish}$ & $15\,\text{m}$ \\
        Beam Size (at $z_\mathrm{eff}=1.675$) & $\theta_\mathrm{FWHM}$ & $2.6\,\text{deg}$ \\
		\hline
	\end{tabular}
    \caption{Specifications for the SKA1-MID Band 1 survey, taken from the SKA Cosmology Red Book \citep{Bacon:2018dui}. 64 of the SKA1-MID dishes will be the existing MeerKAT dishes that have $D_\mathrm{dish}=13.5\,\text{m}$. For simplicity we make the approximation that all dishes have the same diameter as quoted in the table.}
    \label{SKASurveyTable}
\end{table}

\subsection{Simulated Cosmological Signal}\label{CosmoSigSec}

In this work we generate lognormal realizations of the cosmological density field to produce our simulated \hi intensity maps. This provides a computationally efficient way of producing mock data and also provides control of the input cosmology allowing a focused analysis to be carried out on the impacts of 21cm foregrounds. Lognormal density fields were introduced in \citet{ColesLognormal1991} as a way of generating reasonably accurate over-density fields. They have been used in the past by collaborations to generate sets of mock data \citep{Blake:2011wn,Beutler:2011hx} and also for the purposes of \hi intensity mapping \citep{Alonso:2014sna, Cunnington:2020mnn}. 

Beginning with the anisotropic \hi power spectrum as defined by \autoref{eq:PKnofNL}, we obtain the input matter power spectrum $P_\mathrm{m}$ from \texttt{CAMB} \citep{Lewis:1999bs} using a $\Lambda$CDM cosmology based on \citet{Ade:2015xua} where $\{\Omega_\mathrm{m}, \Omega_\mathrm{b}, h, n_\mathrm{s}\} = \{0.307,0.0486,0.677,0.968\}$. The linear growth rate of structure $f$ is approximated by $f \simeq \Omega_\mathrm{m}(z)^\gamma$ where $\gamma$ is the growth rate index \citep{Linder:2005in}. For the linear \hi bias we interpolate values from \citet{Villaescusa-Navarro:2018vsg} to obtain a bias of $b_\hinospace = 1.87$ at our effective redshift $z_\mathrm{eff}=1.675$. Furthermore, for the mean \hi temperature we note that is it related to the \hi density abundance $\Omega_{\hinospace}$, by \citep{Battye:2012tg}
\begin{equation}\label{eq:TbarModelEq}
    \Tbar(z) = 180\Omega_{\hinospace}(z)h\frac{(1+z)^2}{H(z)/H_0} \, {\text{mK}} \, ,
\end{equation}
 where we can use real data constraints from \citet{Masui:2012zc} to set the value for $\Omega_{\hinospace}$ to
\begin{equation}\label{MasuiOmHI}
	\Omega_\hinospace b_\hinospace r = (4.3 \pm 1.1) \times 10^{-4} \,,
\end{equation}
assuming the cross-correlation coefficient $r=1$ and the fiducial \hi bias value.

We follow the lognormal approach in \citet{Beutler:2011hx}, which we outline here for clarity and completeness. By inverse Fourier transforming the power spectrum $P_\hinospace(\bm k)$ we obtain the correlation function $\xi_\hinospace(\bm r)$ which we then transform to $\xi'_\hinospace(r)=\ln(1+\xi_\hinospace(r))$. Applying a Fourier transform to this reverts back to the power spectrum, but it has now been transformed to $P_\hinospace'(\mathbf{}{k})$ such that the lognormal realization of $P_\hinospace'$ will be the same as a Gaussian realization of $P_\hinospace$. We can then follow the standard lognormal procedure \citep{ColesLognormal1991}
\begin{equation}\label{LogNormalEq}
    \delta_\mathrm{LN}=\exp\left(\delta_\mathrm{G}-\frac{\sigma_\mathrm{G}^2}{2}\right) - 1 \, ,
\end{equation}
where $\sigma_\mathrm{G}^2$ is the variance of the Gaussian overdensity field $\delta_\mathrm{G}$, which is generated by producing a Fourier-space Gaussian overdensity field $\tilde{\delta}_\mathrm{G}$ that is a realization of Gaussian random numbers with $\sigma=\sqrt{P'_\hinospace/2N_\mathrm{cell}}$, where $N_\mathrm{cell}$ is the number of cells in the 3D grid. To provide the final lognormal density field of the Gaussian input power spectrum, we transform $\tilde{\delta}_\mathrm{G}$ back to configuration space and follow \autoref{LogNormalEq}.

\subsection{Power Spectrum Measurement}\label{PkMeasurement}

We focus on a Fourier space power spectrum analysis. We embed our simulated data into a cube with lengths $L_\mathrm{x},L_\mathrm{y},L_\mathrm{z}$ which is gridded into $n_\mathrm{x} \times n_\mathrm{y} \times n_\mathrm{z} = 256^3$ cells. As such the volume of each cell is $V_\mathrm{cell} = L_\mathrm{x} \times L_\mathrm{y} \times L_\mathrm{z} / N_\mathrm{cell}$ where $N_\mathrm{cell} = n_\mathrm{x} \times n_\mathrm{y} \times n_\mathrm{z}$. By applying a fast Fourier transform (FFT) to the grid we get the Fourier amplitudes of each mode $\bm k$,
\begin{equation}
	\widetilde{\delta}_\hinospace(\bm k) = \sum_{n=1}^{N_\mathrm{cell}}\delta_\hinospace(\boldsymbol{x}_n) \, \exp({\rm i}\bm k\cdot\boldsymbol{x}_n)\, .
\end{equation}
The SKA1-MID intensity mapping footprint is expected to cover $20,000\,\text{deg}^2$. Converting this to physical dimensions at the effective redshift $z_\mathrm{eff}=1.675$ gives transverse dimensions of $L_\mathrm{x} = L_\mathrm{y} = 8029\,\text{Mpc}/h$. We therefore choose this value for the side lengths of our gridded cube. The SKA Band 1 redshift range of $0.35<z<3$ corresponds to a physical distance of $3458\,\text{Mpc}/h$. We therefore simulate data with this depth in the radial direction then embed that data onto the gridded data cube with dimensions $L_\mathrm{x} = L_\mathrm{y} = L_\mathrm{z} = 8029\,\text{Mpc}/h$. To account for the regions of the cuboid not covered by our data along the radial dimension, we introduce a window function $W(\boldsymbol{x})$ which is 1 where data exists and 0 elsewhere. This is not dissimilar to realistic large-scale structure surveys whose data footprint is typically a light-cone which is embedded onto a Fourier grid and the window function again is 1 inside the cone and 0 outside. From this, the power spectrum estimator is then given as \citep{Blake:2019ddd}
\begin{equation}
	\widehat{P}_\hinospace(\bm k) = \frac{V_\mathrm{cell}\, |\widetilde{\delta}_\hinospace(\bm k)|^2}{\sum_{n=1}^{N_\mathrm{cell}} W^2(\boldsymbol{x}_n)} \, .
\end{equation}
This power spectrum is then spherically averaged into $k$-bins with spacing $\Delta k$ to provide a 1D power spectrum.

The advantage of opting for simulating the density field onto a Cartesian coordinate system is that we can have full confidence behind the plane-parallel approximation. In reality, when surveying scales required to probe PNG, curved-skies need to be taken account which contribute wide-angled effects. This has been investigated and methods exist to deal with such effects \citep{Bianchi:2015oia,Castorina:2017inr,Blake:2018tou}. We therefore make the assumption that curved-sky effects can be controlled when measuring large-scale Fourier modes in intensity maps, but do not need to account for them in our methodology. We note that it is possible to measure clustering on the curved sky directly using harmonic space power spectra. We discuss this further in \secref{Clresults} where we present some foreground-free results using harmonic space $C_\ell$. However, a full investigation would involve a separate analysis and modelling pipeline, along with amendments to our simulations and is therefore something we leave for future work.

In current intensity mapping experiments, errors are dominated by the thermal noise of the instrument \citep{Masui:2012zc, Anderson:2017ert}. However, for the SKA, the thermal noise is expected to be much smaller, therefore our error budget is dominated by cosmic variance, which is of particular importance on the large scales we are interested in. The error on the power spectrum estimator can be written as \citep{Blake:2019ddd}
\begin{equation}
    \sigma_{\widehat{P}}(\bm k) = \frac{1}{\sqrt{N_\mathrm{modes}}} \frac{\widehat{P}_\hinospace(\bm k) + P_\mathrm{N}(\bm k)}{\sqrt{V_\mathrm{frac}}}\, ,
\end{equation}
where $P_\mathrm{N}$ is the power spectrum of thermal (instrumental) noise and discussed further in \secref{InstEffectsSec}, $V_\mathrm{frac}$ is the volume ratio of the SKA1-MID Band 1 survey footprint and the grid size i.e.\ $V_\mathrm{frac} = V_\mathrm{sur}/(L_\mathrm{x}L_\mathrm{y}L_\mathrm{z})$ and $N_\mathrm{modes}$ is the number of unique modes in each $\bm k$-bin and is given by
\begin{equation}
    \frac{1}{N_\mathrm{modes}} = \frac{2(2\pi)^3}{V_\mathrm{sur}\,}\frac{1}{4\pi k^2\Delta k}\, .
\end{equation}

\subsubsection{Theoretical Prediction}
In order to investigate constraints on $\fnl$ and other parameters, we require a model power spectrum to compare to. We use the same \texttt{CAMB} input matter power spectrum $P_\mathrm{m}$ that was used to simulate our cosmological data. To ensure accurate modelling we sample the 1D $P_\mathrm{m}$ over the same 3D Fourier grid dimensions as our data. We then apply the modelling steps discussed in \secref{PNGSec} to account for RSD, bias and intensity mapping systematics as summarised by \autoref{eq:HIPkwithFGbeamModel}, where the input matter power spectrum $P_\mathrm{m}$ now represents the \texttt{CAMB} input power spectrum sampled over our 3D Fourier grid. 

For further consistency with the simulated data, we then convolve this damped model power spectrum with the window function $P_\hinospace(\bm k) \rightarrow P_\hinospace(\bm k) * |W(\bm k)|^2$. Finally, we spherically average this model into the same $k$-bins as used for the data to give our final model 1D power spectrum.

\subsection{Simulating Foreground Contamination}\label{FGSec}
In this work we produce simulated foreground maps from galactic synchrotron, free-free emission (both galactic and extra-galactic) and from point-sources beyond our own Galaxy. We generate realizations of these signals using a power spectrum that forecasts the characteristics of the particular foreground \citep{Santos:2004ju}. For this we use the angular power spectrum $C_\ell$ where scales are defined by multipoles $\ell$ in spherical harmonic space; 
\begin{equation}\label{FGpowerspec}
	C_\ell\left(\nu_1, \nu_2\right)=A\left(\frac{\ell_\mathrm{ref}}{\ell}\right)^\beta\left(\frac{\nu_\mathrm{ref}^2}{\nu_1\,\nu_2}\right)^\alpha \exp \left(-\frac{\log^2\left(\nu_1/\nu_2\right)}{2\,\xi^2}\right) \,.
\end{equation}
\begin{table}
	\centering
	\begin{tabular}{lcccc} 
		\hline
		Foreground & A & $\beta$ & $\alpha$ & $\xi$ \\
        \hline
		Galactic synchrotron & 700 & 2.4 & 2.80 & 4.0\\
		Point sources & 57 & 1.1 & 2.07 & 1.0\\
		Galactic free-free & 0.088 & 3.0 & 2.15 & 35\\
		Extra-galactic free-free & 0.014 & 1.0 & 2.10 & 35\\
		\hline
	\end{tabular}
    \caption{Parameter values for foreground $C_\ell$ (see \autoref{FGpowerspec}) with amplitude $A$ given in mK$^2$. Pivot values used are $\ell_\mathrm{ref} = 1000$ and $\nu_\mathrm{ref} = 130 \, \text{MHz}$ as per \citet{Santos:2004ju}.}
    \label{FGparams}
\end{table}
The values for the fitting parameters are outlined in \autoref{FGparams}. This gives us four separate foreground maps at each frequency and for further completeness we also include a map extrapolated from real data by utilizing the Global Sky Model \citep{deOliveiraCosta:2008pb, Zheng:2016lul}.

For our Band 1 data range of $0.35 < z < 3$, we simulate 109 foreground maps for the corresponding frequency range ($1050\,\text{MHz} > \nu > 350\,\text{MHz}$) giving each a frequency width of $\Delta\nu\sim 6\,\text{MHz}$. The foreground maps, which are output as \texttt{HEALPix} \citep{2005ApJ...622..759G,Zonca2019} maps, are then approximately flattened into Cartesian coordinate maps. Whilst this will cause some angular distortions to the foreground map, the angular coherence of the foregrounds is not something we are overly interested in since this should be largely removed in a successful foreground clean. As with the simulated \hi data, the foreground data is then embedded into the $N_\mathrm{cell} = 256^3$ gridded cube. For further details on this process we refer the reader to \citet{Cunnington:2020mnn}, where an identical method was adopted and discussed in more detail. 

Once these foreground maps are added onto our \hi intensity maps, we then need a process for removing them, allowing us to investigate the effects a foreground clean will have and whether a precise and accurate measurement of $\fnl$ is possible. In this work we use Principal Component Analysis (PCA) and refer the reader to \citet{Alonso:2014dhk} for a more detailed discussion on the approach. To summarise, the method relies on the fact that because the maps should be very correlated through frequency, the majority of the information will be contained in a small number of very large eigenvalues in the frequency covariance matrix. By computing this covariance matrix for the observed data, we can then identify the largest eigenvalues for the system which we remove from the data. This removes the most frequency correlated information, which is assumed to contain most of the 21cm foregrounds.

\begin{figure}
	\centering
  	\includegraphics[width=\columnwidth]{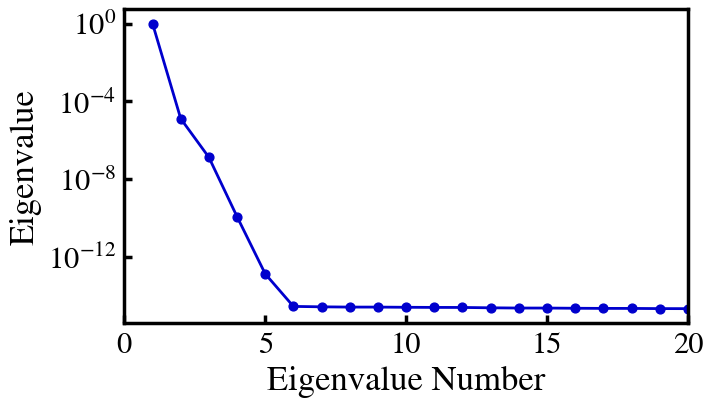}
    \caption{Eigenvalues of the frequency covariance matrix for one simulated SKA1-MID Band 1 data-set ordered by magnitude and normalized such that all eigenvalues sum to unity. This suggests that the vast majority of the frequency correlated signal is contained within the first 5 eigenvalues.}
\label{fig:PCAeigenval}
\end{figure}

We demonstrate this process in \autoref{fig:PCAeigenval}, where we plot the eigenvalues for the frequency covariance matrix for a lognormal realized \hi intensity map with foreground contamination. We order the eigenvalues by size and it is clear from the plot that there are 5 eigenvalues that dominate the rest and it is within these that the frequency correlated modes should exist.

\autoref{fig:ConstituentPks} shows the power spectra for the simulated data decomposed into various contributions. This firstly shows the dominance the foregrounds (contained in the \textit{uncleaned} observed \hi emission - purple solid line with crosses) have over the \hi cosmological signal (black solid line). However, by removing $N=5$ eigenvalues (pink thin line with circles), we can remove most of this contamination and largely recover the \hi signal. We also show some other examples of different numbers of modes removed to demonstrate how this affects the clean. It is evident that using $N=3$ or 4 is not sufficient. It also shows that going to $N=6$ provides no discernible improvement to the clean. We therefore chose to use $N=5$ in our main forecasts. We also include contributions from the instrumental noise and shot noise as the grey lines (dashed and dotted respectively). These are discussed in the following \secref{InstEffectsSec}.

One final point to highlight from \autoref{fig:ConstituentPks} is that 
the $N=5$ PCA clean is marginally below the original \hi signal at small-$k$. This means the foreground clean is damping power on the largest scales and this is a well-known consequence of foreground cleaning. This has the potential to be a problem for probing PNG with intensity mapping and will be the main focus of our investigation.

\subsection{Instrumental Effects}\label{InstEffectsSec}

\subsubsection{Thermal Noise}\label{ThermalNoise}

\begin{figure}
	\centering
  	\includegraphics[width=\columnwidth]{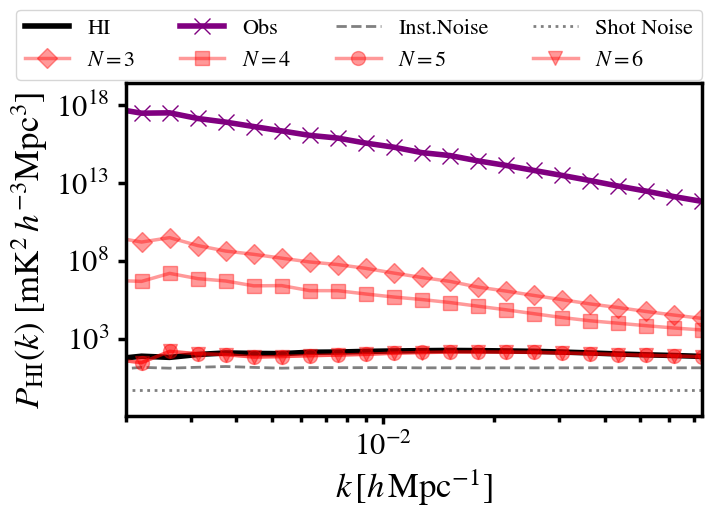}
    \caption{Impact of foregrounds on the power spectrum and performance from PCA cleaning. \textit{Black-solid} line is the \hinospace-signal-only power and the \textit{purple-crossed} line shows the overall observed \hi emission which is dominated by foregrounds. The \textit{pink} lines show a PCA clean with differing number of modes removed, indicated by the different markers, we use $N=5$ in all other results presented. Lastly, the \textit{grey-dashed} and \textit{grey-dotted} lines show the instrumental and shot noise components.}
\label{fig:ConstituentPks}
\end{figure}

The system temperature for the SKA1-MID instrument is  given by  \citep{Bacon:2018dui}
\begin{equation}
\begin{split}
    T_\mathrm{sys} = 15\text{K} + \ & 30\text{K} \left(\frac{\nu}{\text{GHz}} - 0.75\right)^2 \\
    \ & + T_\mathrm{spl} + T_\mathrm{CMB} + 25\text{K}\left(\frac{408\,\text{MHz}}{\nu}\right)^{2.75} \, ,
\end{split}
\end{equation}
where $T_\mathrm{spl}\simeq 3\,\text{K}$ and $T_\mathrm{CMB}\simeq 2.73\,\text{K}$. These specifications can be used to model instrumental noise as uncorrelated white noise. We generate a Gaussian random field with 
\begin{equation}\label{eq:Noise}
    \sigma_\mathrm{noise} = T_\mathrm{sys} \sqrt{\frac{4\pi \, f_\mathrm{sky}}{\Omega_\mathrm{beam} \, N_\mathrm{dish} \,  t_\mathrm{obs} \, \delta\nu}}\, ,
\end{equation}
where $\Omega_\mathrm{beam} \simeq 1.133\, \theta_{\mathrm{FWHM}}^{2}$ is the solid angle for the instrumental beam and the rest of the parameters are defined in \autoref{SKASurveyTable}. Due to the large number of dishes, large amount of observation time, and an expected low system temperature, an SKA1-MID single-dish intensity mapping survey is expected to have a reasonably low level of instrumental noise ($\sigma_\mathrm{noise}\simeq0.021\,\text{mK}$ at $z_\mathrm{eff}=1.675$) and is not expected to dominate over the cosmological \hi signal ($\sigma_\hinospace\simeq0.036\,\text{mK}$). This is shown in \autoref{fig:ConstituentPks} by the grey dashed line.

\subsubsection{Radio Telescope Beam}
We also aim to approximate the effects of the radio telescope beam, which as we have seen is quite large for a single-dish SKA intensity mapping survey. As an example, in \autoref{SKASurveyTable} we quote the beam size at the Band 1 central redshift where the beam is expected to be $\theta_\mathrm{FWHM}\simeq2.62\,\text{deg}$. This is calculated from $\theta_\mathrm{FWHM}=1.22\,\lambda_\mathrm{21cm}(1+z)/D_\mathrm{dish}$.

In this investigation we are targeting the largest scales possible and it is therefore unlikely the beam will have significant impact. A beam of the size quoted in \autoref{SKASurveyTable} equates to a physical scale of $63\,\text{Mpc}/h$, which is a much smaller scale (higher-$k$) than where we would expect a scale-dependent bias to be observable. Despite this we still simulate the beam by convolving the map ($\delta T_\hinospace + \delta T_\mathrm{FG} + \delta T_\mathrm{noise}$) with a symmetric, two-dimensional Gaussian function with a full-width-half-maximum of $\theta_\mathrm{FWHM}$ acting in the direction perpendicular to the LoS. In our modelling we assume perfect knowledge of this beam size (\secref{ModellingIMsystematics}) and, as with other parameters, we assume it is non-evolving through frequency (note that this would not be the case for a real survey and a frequency-dependent beam would need to be accounted for along with other evolving parameters).

\subsubsection{Shot-Noise}
For resolved galaxy surveys the shot-noise (or Poisson noise) can be quite a dominating component and goes as $P_\mathrm{SN}=1/N_\mathrm{g}$ where $N_\mathrm{g}$ is the number of galaxies detected in the survey. For \hi intensity mapping, whilst the shot-noise is not yet fully understood \citep{Villaescusa-Navarro:2018vsg, Spinelli:2019smg}, it is largely agreed that its contribution will be minimal owing to the fact that the signal obtained is integrated over all galaxies down to the faintest. We show an approximate shot-noise amplitude in the context of the other signal contributions in \autoref{fig:ConstituentPks}. This value is interpolated from hydrodynamical simulations performed in \citep{Villaescusa-Navarro:2018vsg}. As can be seen, the shot-noise contribution is very subdominant and as such we do not attempt to include it in our simulated data nor model it in these forecasts.

\section{MCMC forecasts}\label{MCMCresults}
Here we perform Bayesian MCMC analyses on our data, to see how well we can recover the true, input (fiducial) $\fnl = 0$ value of our simulations, and how well we can constrain it. Other parameters for the model are outlined in \autoref{tab:fidParams}. Given that the simulated data has been produced from the same input \texttt{CAMB} matter power spectrum that we are comparing to in the model, we expect to recover this and any other input fiducial parameters.

In addition, since we are simulating instrumental noise, telescope beams and covering a range of scales representative of an SKA1-MID Band 1 survey, the error on the constraints obtained should be a representative forecast for this survey. Furthermore, when we analyse the effect of foregrounds, the constraints on $\fnl$ become more dependent on our adopted foreground model, and serve as a robust test of measuring $\fnl$ in the presence of 21cm foregrounds. We utilize the public code \texttt{emcee}\footnote{\href{https://emcee.readthedocs.io/en/stable/}{emcee.readthedocs.io}} \citep{ForemanMackey:2012ig} to explore parameter space using an MCMC method where in the log-likelihood, we use our model from \autoref{eq:HIPkwithFGbeamModel} and in each case we use 200 walkers and 800 samples.
\begingroup
\setlength{\tabcolsep}{10pt} 
\renewcommand{\arraystretch}{1.2} 
\begin{table}
	\centering
	\begin{tabular}{lcc} 
		\multicolumn{3}{c}{\textbf{Model Parameters}} \\
		\hline
		Parameter & Fiducial Value & Prior \\
		\hline
		$\fnl$ & 0 & $-6.0 < \fnl < 4.2$ \\
		$\Tbar$ [mK] & $0.0781$ & $0.02 < \Tbar < 0.15$ \\
		$k^\text{FG}_\parallel$ [$h\,\text{Mpc}^{-1}$] & $3.1\times 10^{-3}$ & $10^{-3} < k^\text{FG}_\parallel < 10^{-2}$ \\
		\hline
	\end{tabular}
    \caption{Assumed fiducial parameters for our model. We assume flat priors in all cases. The $\fnl$ \citet{Akrami:2019izv} prior is only used in cases where it is explicitly stated, otherwise we assume a very wide prior of $-150<\fnl<150$. The other parameters are evaluated at the central effective redshift of SKA1-MID Band1 survey ($z_\mathrm{eff}=1.675$). The fiducial value of $\kFG$ is a fit by eye to our data in \autoref{fig:ModelledFullPks} with $\fnl=0$. The priors for $\Tbar$ and $k^\text{FG}_\parallel$ are used for cases where we allow these parameters to vary. The other parameters in our model are fixed throughout and, unless otherwise stated,  they are $b_\hinospace = 1.87$, $f=0.941$, and $\alpha_\mathrm{FG}=0.97$.} 
    \label{tab:fidParams}
\end{table}

\autoref{fig:ModelledFullPks} shows the averaged measured power spectra from 100 simulated datasets, following the log-normal plus systematic simulations we presented in \secref{Methodology}. The black crossed markers represent the foreground-free \hi intensity maps and the red circled markers represent the same \hi intensity maps but with foregrounds cleaned using PCA with $N=5$. We also include the model predictions outlined by \autoref{eq:HIPkwithFGbeamModel}. The red dotted-line includes damping from the foregrounds as modelled by \autoref{eq:FGmodel} and we use the fixed value of $k^\text{FG}_\parallel = 3.1\times 10^{-3}\,h\,\text{Mpc}^{-1}$, which has been fitted for. The black dashed-line is the foreground-free model equivalent to setting $\widetilde{B}_\mathrm{FG}=1$ in \autoref{eq:HIPkwithFGbeamModel}. We can see an excellent agreement between simulated data and model. This is expected for the no foreground case where we are modelling the power spectrum with the same input power spectrum used in the log-normal simulation. However, in the subtracted foreground case, we are truly modelling a state-of-the art foreground removal procedure with a high level of accuracy -- with the caveat that this is still a fairly idealised situation due to absent real data complications that will be discussed later on.
\begin{figure}
	\centering
  	\includegraphics[width=\columnwidth]{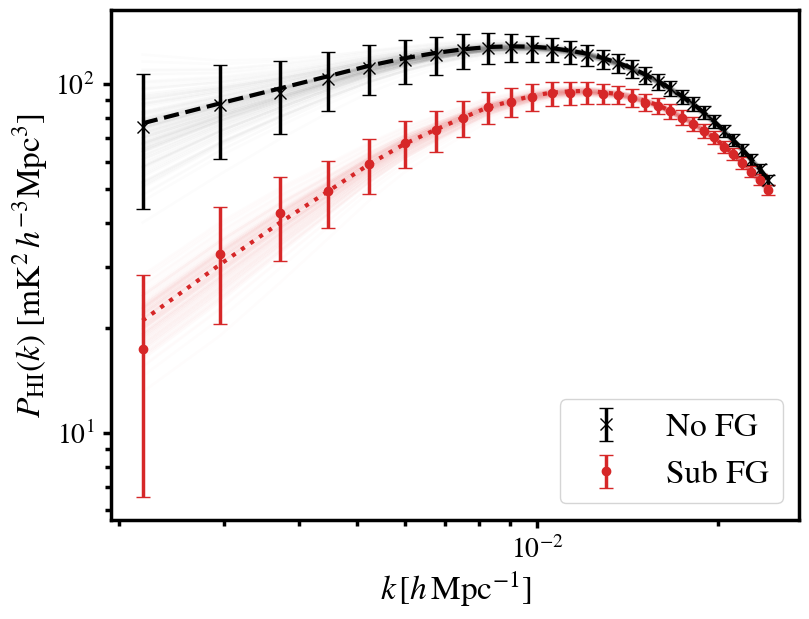}
    \caption{Measured power spectra from simulated SKA1-MID Band 1 data. \textit{Black-crossed} markers represent foreground-free data and \textit{red-circled} markers are for maps with foregrounds included and cleaned using PCA. \textit{Dotted-red} line shows the model represented by \autoref{eq:FGmodel} and the \textit{dashed-black} line shows the model with no damping from foregrounds i.e.\ $\widetilde{B}_\mathrm{FG}=1$. The thin lines show some samples from the MCMC.}
    \label{fig:ModelledFullPks}
\end{figure}

For the results presented by \autoref{fig:ModelledFullPks}, where we assume all other parameters fixed, we can recover a constraint of $\fnl = -1.38_{-3.16}^{+3.56}$ (68\% CL) in the foreground-free case with a flat prior on the $1\sigma$ range of the \citet{Akrami:2019izv} results. Interestingly, by including foregrounds with a fixed model for their contamination, the constraint does not seem to be affected, but again this is while using the \citet{Akrami:2019izv} prior. In \autoref{tab:fNLsummary}, we summarize our recovered constraints on $\fnl$ using an MCMC under different parameter varying and prior choices, which we present in this section. Removing the \citet{Akrami:2019izv} prior weakens the constraints as expected, but we still obtain competitive bounds of $\fnl = -3.98_{-9.41}^{+8.80}$ (no foregrounds) and $\fnl = -2.94_{-11.9}^{+11.4}$ (with foregrounds).
\begingroup
\setlength{\tabcolsep}{10pt} 
\renewcommand{\arraystretch}{1.6} 
\begin{table}
	\centering
	\begin{tabular}{lc}
        \multicolumn{2}{c}{$f_\mathrm{NL}$ \textsc{Constraints with 21cm Intensity Mapping}} \\
        \specialrule{.2em}{.1em}{.1em}
        
        \multicolumn{2}{c}{\textbf{Foreground-Free}} \\
        \specialrule{.2em}{.1em}{.1em} 
        Method &
        $f_\mathrm{NL}$ \\
        \hline
		Fixed parameters (with Planck18 prior) & $-1.38_{-3.16}^{+3.56}$ \\
		Fixed parameters & $-3.98_{-9.41}^{+8.80} $ \\
		Varying $\Tbar$ & $-3.58_{-12.8}^{+11.8}$ \\
		Harmonic Space $C_\ell$ - Varying $\Tbar$ & $-2.35_{-10.4}^{+7.60}$ \\
		& \\
        \specialrule{.2em}{.1em}{.1em} 
        
        \multicolumn{2}{c}{\textbf{Subtracted Foregrounds}} \\
        \specialrule{.2em}{.1em}{.1em} 
        Method & $f_\mathrm{NL}$ \\
        \hline
		Fixed parameters (with Planck 18 prior) & $-1.11_{-3.25}^{+3.48}$ \\
        Fixed parameters & $-2.94_{-11.9}^{+11.4}$ \\
		Varying $\Tbar$ & $-0.46_{-18.7} ^{+17.9}$ \\
		Varying $k^\text{FG}_\parallel$ & $+0.75_{-44.5}^{+40.2}$\\
		Varying $k^\text{FG}_\parallel$ (with Planck 18 prior) & $-0.74_{-3.55}^{+3.42}$ \\
        \specialrule{.2em}{.1em}{.1em} 
        
	\end{tabular}
    \caption{Summary of recovered values for $\fnl$ with 68\% confidence intervals under various methods. All results are representative of an SKA1-MID Band 1 intensity mapping survey. \textit{Top}-table shows the foreground-free cases and the \textit{bottom}-table is where foregrounds have been added, then cleaned using PCA and modelled using \autoref{eq:FGmodel}. Fixed parameters refers to the case where all parameters except $\fnl$ in our model have been fixed to their fiducial values (summarized in \autoref{tab:fidParams}). The Planck18 prior refers to $\fnl = -0.9 \pm 5.1$ \citep{Akrami:2019izv}, all other cases have no prior on $\fnl$.} 
    \label{tab:fNLsummary}
\end{table}

The recovered values outlined in \autoref{tab:fNLsummary} showcase the success of the foreground contamination modelling we are using. The parameters in the \emph{subtracted foregrounds} part of the table are unbiased with varying amounts of uncertainty added into the constraint when compared to their \emph{foreground-free} counterparts. In \autoref{fig:CornerNoFGModel}, we further justify the need for a foreground model by comparing it to a situation where one is not used. This shows the MCMC results where we allow $\Tbar$ to vary along with $\fnl$. The red-solid contour demonstrates that when using our foreground model, an unbiased recovery of parameters is obtained. The purple-solid contour and purple dotted-lines show the result where $B_\mathrm{FG}=1$, equivalent to using no foreground model. The heavily biased result is motivation for the requirement of either a foreground model (such as we have used) or a trusted foreground transfer function (as previously discussed in \secref{ModellingIMsystematics}).
\begin{figure}
	\centering
  	\includegraphics[width=\columnwidth]{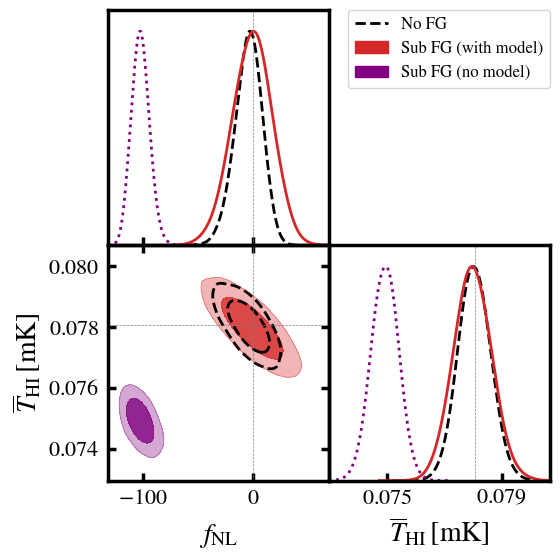}
    \caption{Demonstrating the requirement for a model of foreground contamination. Contours represent 68\% and 95\% confidence regions. \textit{Purple-solid} contour and dotted lines show the biased results obtained for both $\fnl$ and $\Tbar$ where a model is not used on foreground cleaned data i.e.\ by setting $\widetilde{B}_\mathrm{FG}=1$. \textit{Red-solid} contour and solid line show the same case but implementing the foreground model as per \autoref{eq:FGmodel}. Modelled foreground results agree with the foreground-free case (\textit{black-dashed}), just with slightly higher uncertainties due to the requirement of extra model parameters.}
    \label{fig:CornerNoFGModel}
\end{figure}

\subsection{Neutral Hydrogen Abundance}
An unknown quantity relating to the \hi power spectrum is the mean \hi temperature $\Tbar$. In our modelling, we are able to set this to the known-value used in our simulations; this is, however, not the case with real data. The uncertainty in this parameter comes from uncertainty of the \hi abundance and the two are proportional ($\Tbar \propto \Omega_\hinospace$) as shown in \autoref{eq:TbarModelEq}. Whilst the \hi density abundance $\Omega_\hinospace$ has been relatively well constrained at low redshifts ($z<0.1$) with targeted \hi galaxy surveys \citep[e.g.][]{Rhee:2013fma} and at higher redshifts ($z>2$) with Lyman-$\alpha$ probes \citep[e.g.][]{Noterdaeme:2012gi}, at mid-redshifts it is poorly known.

Therefore, it is more realistic to let $\Tbar$ vary, as done for the confidence contours obtained in \autoref{fig:CornerNoFGModel}. As summarised in \autoref{tab:fNLsummary}, letting $\Tbar$ vary between some wide prior (given by \autoref{tab:fidParams}) degrades the constraints compared to the fixed $\Tbar$ case to $\fnl = -3.58_{-12.8}^{+11.8}$ (no foregrounds) and $-0.46_{-18.7}^{+17.9}$ (with foregrounds). Note that allowing $\Tbar$ to vary in the MCMC analysis is identical to including the overall amplitude of the power spectrum as a free parameter. In absence of RSD this is then also akin to allowing the Gaussian bias $b_\hinospace$ to jointly vary with  $\Tbar$. There is well known degeneracy between $\Tbar$ and $b_\hinospace$ but it is hoped future surveys may be able to lift this degeneracy by using RSD \citep{Masui:2012zc,Pourtsidou:2016dzn}. For simplicity, in this work we choose not to address this degeneracy and take the approach of just varying the overall amplitude, i.e.\ $\Tbar$, and leave $b_\hinospace$ fixed. A similar approach was used in \citet{Villaescusa-Navarro:2016kbz} where a parameter $b_\mathrm{fit} \propto b_\hinospace\Omega_\hinospace$ was used and interpreted as the overall bias of the 21cm signal.

For the foreground-removed case, the parameter $\alpha_\mathrm{FG}$ in our model (see \autoref{eq:FGmodel}) is required to avoid obtaining biased $\Tbar$ results; $\alpha_\mathrm{FG}$ acts as a scaling factor to slightly damp all modes due to the fact that contamination from foreground removal is not perfectly isolated to large radial modes. We demonstrated in \autoref{fig:TransfervsBfg} the success of this model and in \autoref{fig:alpha=1plot} we further motivate the inclusion of $\alpha_\mathrm{FG}$ by showing results where this parameter is removed (by setting $\alpha_\mathrm{FG} = 1$). This shows that in the foreground removed case (red contours), the recovered value on $\Tbar$ is biased and outside the 95\% confidence region. Interestingly, this does not seem to bias the estimation of $\fnl$. We fix  $\alpha_\mathrm{FG} = 0.97$ for all our remaining results.

The contours in both \autoref{fig:CornerNoFGModel} and \autoref{fig:alpha=1plot} suggest an anti-correlation exists between $\Tbar$ and $\fnl$. This can be understood by considering the effect of a positive $\fnl$ is to boost power at small-$k$ due to the addition of a positive scale-dependent bias (see \autoref{fig:ModelfNL} for demonstration). In response to this enhancement of power from positive $\fnl$, the amplitude of the power spectrum must decrease to compensate, hence $\Tbar$ must decrease. 
\begin{figure}
	\centering
  	\includegraphics[width=\columnwidth]{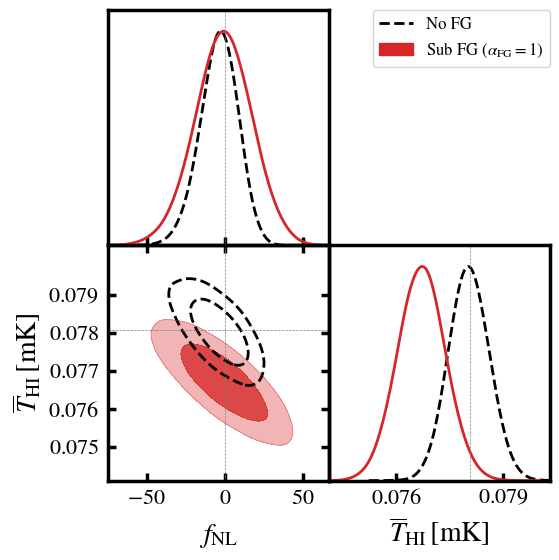}
    \caption{Effect of excluding the $\alpha_\mathrm{FG}$ parameter from the foreground model. This shows 68\% and 95\% confidence regions for parameters $\fnl$ and the mean \hi temperature $\Tbar$ with all other parameters fixed but with $\alpha_\mathrm{FG}=1$. Omitting the scaling parameter $\alpha_\mathrm{FG}$ does not bias $\fnl$ but causes a bias in $\Tbar$ (red-solid contour).}
    \label{fig:alpha=1plot}
\end{figure}

\subsection{Degeneracies with a Nuisance Foreground Model}

The demonstration of successful $\fnl$ recovery so far, has come under scenarios where we tune the foreground parameter $k_\parallel^\text{FG}$ and fix it in the MCMC analysis. It is not clear that this approach will be viable with real SKA data as we may still lack an understanding of foreground contamination to be able to proceed with such rigid models. It is more likely that various parameters in the foreground model will need to be varied as nuisance parameters and marginalised over. We investigate this scenario in \autoref{fig:Corner_fNL_kFG} where we vary $\fnl$ along with the model parameter $\kFG$, leaving all other parameters fixed. The degenerate contour produced here suggests that a precise measurement of $\fnl$ using intensity mapping is intrinsically linked to how well foreground contamination can be understood. Treating $\kFG$ as a nuisance parameter and marginalising over it severely hinders constraints to $\fnl = 0.75_{-44.5}^{+40.2}$. The degeneracy we find  can be analytically approximated by $\fnl = 53\times 10^3(\kFG - 3.1\times 10^{-3})$, but we stress that this not generalizable and will depend upon a survey's dimensions and the instrument's response to foregrounds.

\autoref{fig:Corner_fNL_kFG} therefore highlights a key conclusion from this paper -- that a degeneracy exists between estimates and constraints on $\fnl$ and the modelling of foreground contamination. This means that a non-zero $\fnl$ in real data could be misconstrued as poor fit of foreground contamination.

\begin{figure}
	\centering
  	\includegraphics[width=\columnwidth]{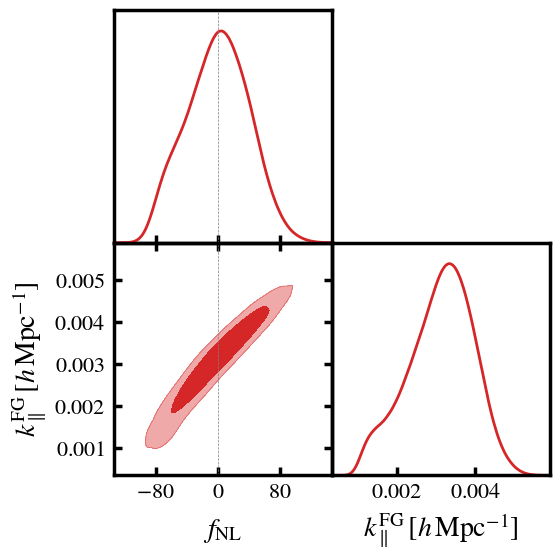}
    \caption{Degenerate contour between $\fnl$ and $\kFG$ (the key parameter in our model of foreground contamination). Contours represent 68\% and 95\% confidence regions. This indicates that constraints on $\fnl$ are highly dependent on how well foreground contamination can be modelled. This scenario gives a disappointing constraint of $\fnl = 0.75_{-44.5}^{+40.2}$ compared to the case where we fix the free parameter $\kFG$ ($\fnl = -2.94_{-11.9}^{+11.4}$ -  \autoref{tab:fNLsummary}).}
    \label{fig:Corner_fNL_kFG}
\end{figure}

\subsection{Harmonic-Space Power Spectrum}\label{Clresults}
In this section, we comment on the possibility to probe a scale-dependent bias in harmonic space using the \hi angular power spectrum $C^\hinospace_\ell$ \citep{Camera:2014bwa,Fonseca:2015laa}. We conducted a small investigation on this approach using the same SKA1-MID Band 1 assumptions and found a constraint of $\fnl = -2.35_{-10.4}^{+7.60}$ in the absence of foregrounds. Testing this required a completely different simulation setup and modelling process, which we outline in \appref{HarmonicAppendix}. Comparing this constraint with the analogous scenarios in \autoref{tab:fNLsummary} shows this approach achieves similar estimates, as expected (slightly lower uncertainties but same fractional errors). The reason we have not investigated this approach further by including the effects of foregrounds is due to simulation complexities. In order to add frequency binned foreground emission over the SKA Band 1 range, we would also need to obtain coherent spherical shells of \hi intensity mapping data. Simulating this would require a higher resolution log-normal realization of the density field in a cube, then sampling this cube to extract spherical shells which would represent \texttt{HEALPix} maps at different frequencies. Previous work has carried out simulations similar to this \citep[e.g.][]{Alonso:2014sna, Witzemann:2018cdx}, but we leave this further investigation for future work.

It is important to comment on the potential of this alternative methodology. An attractive feature of angular power spectra is that they are a more direct observable since they do not need the assumption of a cosmology to convert angles to distances and they are also gauge-independent, potentially important for probing $\fnl$ \citep{Bruni:2011ta,Jeong:2011as}. In the case of \hi intensity mapping, measuring modes in harmonic space may deliver a different response from foreground contamination, and it would be interesting to investigate how easy it is to model. This is ample motivation for future study and it would be interesting to robustly compare the two approaches including foregrounds.

\section{Discussion}\label{Discussion}

All results summarized in \autoref{tab:fNLsummary} are unbiased constraints on $\fnl$ within 68\% confidence. However, despite averaging over 100 simulations, the majority of our $\fnl$ constraints are slightly negative (below the $\fnl=0$ fiducial input). We found that the measured power spectra of our simulated data was consistently $\sim99.5\%$ of the input power spectrum from which the realization was generated. This is a well documented feature for the simulation approach we adopt \citep{Xavier:2016elr,Agrawal:2017khv} and we believe this is the reason for the consistently negative recovered values of $\fnl$. However, since our MCMC analysis avoided returning biased confidence regions, this is a very minor effect.

Even putting aside the degeneracy problem between $\fnl$ and foregrounds (\autoref{fig:Corner_fNL_kFG}), the constraints we find perhaps present a more pessimistic case for probing PNG with intensity mapping in comparison to the $\sigma(\fnl)\sim 1$ forecasts previously found \citep{Camera:2013kpa,Alonso:2015uua,Fonseca:2015laa}. It is worth emphasizing the different approaches these studies have taken in comparison to the one we have presented here. These studies use a Fisher matrix analysis to compute their forecasts. In this work, we have produced simulated data sets which require a measurement and modelling pipeline in order to recover predicted values of $\fnl$ that serve as forecasts. The optimism of such forecasts are therefore linked to the efficacy of the pipeline and we do not claim to have constructed a maximally optimal pipeline for probing PNG with intensity mapping, nor have we aimed to. This is of course possible, and would hopefully provide improved constraints, but in this work we concentrate on the foreground cleaning implications for probing PNG. Furthermore, the Fisher matrix forecasts, by definition, return the best possible constraints under certain survey assumptions and it is therefore expected that they will always represent the most optimistic scenario.

Further confidence in forecasts can only come with improved understanding of other parameters linked to the surveying technique. For example, we found a degradation in $\fnl$ constraints when we allowed $\Tbar$ to vary within a large conservative prior (\autoref{tab:fidParams}). It is hoped that, by the time SKA1 comes to undertake the large-scale surveys required to probe PNG, the understanding of parameters like the \hi abundance will be more controlled and stricter priors can be used, in turn improving $\fnl$ constraints. Pathfinder surveys such as MeerKAT \citep{Santos:2017qgq,Pourtsidou:2017era} will also shape our understanding of \hi intensity mapping, particularly in terms of foreground contamination. So far, successful \hi intensity mapping detections have entirely relied on cross-correlation with other surveys where uncorrelated residual foregrounds and other systematics likely drop out from the power spectrum measurement \cite{Masui:2012zc,Anderson:2017ert}. Whilst this is a great success, less is learnt about the foregrounds this way compared with an auto-correlation approach where a detailed understanding of foreground contamination and other systematics is required. 

It is not generally known how many components in the source separation will need to be removed from intensity maps to clean foregrounds. In this work we used $N=5$ PCA modes (see \autoref{fig:ConstituentPks}), but this tends to be simulation specific and can be much higher as has been the case with pathfinder intensity mapping surveys \citep[e.g.][]{Wolz:2015lwa}. Furthermore, if the effects of polarization leakage are included a more aggressive clean is required \citep{Carucci:2020enz}. We have not included polarization leakage in this study and assumed an SKA1-MID survey will have sufficient control over this, but it is unclear if the high calibration requirements will be met to avoid this in a single-dish approach. A higher number of modes removed in a foreground clean will inevitably affect the model we have developed and potentially the degeneracies we have identified. Hence, further investigation is needed to see how our model responds under differing foreground cleaning requirements.

We have taken the approach of ensuring little foreground residual remains in our simulated data. This invariably causes a damping of \hi power (\autoref{fig:ModelledFullPks}) due to \textit{over}-cleaning. An alternative approach is to adopt a less aggressive clean and accept some foreground residual will be present in the \hi intensity map, due to \textit{under}-cleaning. This would require a slightly different modelling approach to what we have presented and would also increase the error budget, weakening the constraints. Cross-correlating under-cleaned \hi intensity maps with e.g.\ galaxy surveys, is a way to cause the foreground residuals to drop out the power spectrum whilst the less aggressively cleaned \hi signal correlates with the alternative tracer.  A very popular method to probe PNG is using the multi-tracer approach, which is hugely beneficial due to cosmic variance cancellation \citep{Fonseca:2015laa,Alonso:2015sfa,Ballardini:2019wxj}. The most appealing large-scale surveys to cross-correlate are SKA intensity maps with a next-generation photometric optical galaxy catalogue, see e.g. \citet{Witzemann:2018cdx,Cunnington:2019lvb,Padmanabhan:2019xhc}. This makes the multi-tracer approach particularly important, but the issues we mentioned above remain: over-cleaning introduces a bias, while under-cleaning introduces an additional error, reducing the effectiveness and applicability of the aforementioned techniques \citep{Witzemann:2018cdx}. 

For probing scales beyond the Hubble horizon, potential relativistic effects \citep{Camera:2014sba,Fonseca:2015laa,Wang:2020ibf} would need to be considered which also introduce scale-dependent corrections, thus introducing a further degeneracy. Our analysis has not included such considerations but an extended investigation would inevitably show that relativistic effects would increase uncertainties in $\fnl$ or potentially bias results if not sufficiently considered. Finally, we also note that future experiments probing PNG will likely need to account for likelihood non-Gaussianity in order to obtain accurate $\fnl$ estimates \citep{Hahn:2018zja}.

\section{Conclusion}\label{Conclusion}
In this work we simulated large-scale \hi intensity mapping data sets for an SKA1-MID-like single-dish intensity mapping survey. Crucially, this included added maps of 21cm foreground emission, which we cleaned using techniques likely to be used on real data. This allowed us, for the first time, to investigate probing PNG with realistic effects from 21cm foreground contamination considered. The results presented suggest a worrying degeneracy exists between the PNG parameter $\fnl$, and the modelling of the contamination from foreground cleaning.\newline
\\
\noindent We summarize our main conclusions below:

\begin{itemize}[leftmargin=*]

\item The degeneracy shown in \autoref{fig:Corner_fNL_kFG} between $\fnl$ and $\kFG$ (the main parameter in our model of foreground contamination) suggests constraining PNG is largely dependent on how well effects from foreground cleaning can be understood. If we can confidently apply a fixed model to the damping of power from the foreground clean, then reasonable constraints of $\fnl = -2.94_{-11.9}^{+11.4}$ can be achieved. However, if parameters in the model need to be varied as nuisance parameters and marginalized over (as was the scenario for \autoref{fig:Corner_fNL_kFG}), constraints degrade to $\fnl = 0.75_{-44.5}^{+40.2}$.
\\
\item The above results assume no prior information on $\fnl$ but we also explored using priors from the independent \citet{Akrami:2019izv} data. As one would expect this greatly improves the constraints to $\fnl = -0.74_{-3.55}^{+3.42}$.
\\
\item The approach of this study demanded the construction of a working model that addresses the effects of a foreground clean (motivated by \autoref{fig:CornerNoFGModel}), which allowed an exploration into how degenerate such a model is with $\fnl$. A
welcomed by-product from this investigation is therefore our foreground model (\autoref{eq:FGmodel}). The model includes two free parameters; $\kFG$ -- the tunable parallel wave-vector scale below which signal is highly damped, and $\alpha_\mathrm{FG}$ -- a small ($\sim0.97$) scalar damping factor to account for power damping across all scales. We argued that the $k^{\min}_\parallel$ parameter in the Heaviside part of the model (\autoref{eq:Heaviside}) should not be a free parameter and is simply the smallest parallel scale accessible by the given survey, which is removed since these will be entirely foreground dominated even in the most optimistic scenarios.
\\
\item We carried out some basic tests for our foreground model in \autoref{fig:TransfervsBfg} that largely demonstrated a good performance. Our later results focused heavily on the $\kFG$ parameter, which highlighted its sensitivity to large modes. We also showed the effect of changing $\alpha_\mathrm{FG}$ which  we found did not bias $\fnl$ but did bias $\Tbar$ (\autoref{fig:alpha=1plot}). This could lead to additional complications for constraining $\Omega_\hinospace$ using foreground contaminated intensity maps. Clearly, further work is needed to explore how generally this model can be applied for differing survey sizes and under differing levels of foreground cleaning requirements.
\\
\item Our analysis has mainly relied on simulations onto a Cartesian Fourier grid allowing a perfect plane-parallel approximation to be made. In future work, more realistic curved skies can be simulated and embedded onto the Fourier grid with a treatment of wide-angle effects. Furthermore, our harmonic-space investigation can be extended under this regime with foregrounds included. This will allow a more robust comparison of the $P(k)$ and  $C_\ell$ methods. Our preliminary results into this 
found similar constraints, albeit in the absence of foregrounds and with two largely differing simulation approaches -- we leave further investigation of this method and a direct comparison with $P(k)$ for future work.
\\
\item The forecasts we have presented come with the caveat that they have relied on a pipeline that could be further optimized for the task of measuring $\fnl$. This is something other works have placed large investment into (e.g.\ \citet{Mueller:2017pop, Castorina:2019wmr}). However, our work has clearly demonstrated the importance of placing more consideration to the 21cm foreground problem when probing ultra-large scales with \hi intensity mapping.

\end{itemize} 

\section*{Acknowledgements}
We are grateful to Chris Blake, Hamsa Padmanabhan and Mario Santos for useful comments. SCu is supported by STFC grant ST/S000437/1. AP is a UK Research and Innovation Future Leaders Fellow, grant MR/S016066/1, and also acknowledges support by STFC grant ST/S000437/1. SCa and AP are grateful to the Royal Society International Exchanges grant IES$\setminus$R1$\setminus$180189, which initiated this project. SCa acknowledges support from the Italian Ministry of Education, University and Research (\textsc{miur}) through the `Departments of Excellence 2018-2022' Grant (L.\ 232/2016) awarded by \textsc{miur} and Rita Levi Montalcini project `\textsc{prometheus} -- Probing and Relating Observables with Multi-wavelength Experiments To Help Enlightening the Universe's Structure', in the early stages of this project. This research utilised Queen Mary's Apocrita HPC facility, supported by QMUL Research-IT \url{http://doi.org/10.5281/zenodo.438045}. We acknowledge the use of open source software \citep{scipy:2001,Hunter:2007,  mckinney-proc-scipy-2010, numpy:2011}. Some of the results in this paper have been derived using the \texttt{healpy} and \texttt{HEALPix} package.

\section*{Data Availability}

The data underlying this article will be shared on reasonable request to the corresponding author.




\bibliographystyle{mnras}
\bibliography{Bib} 



\appendix

\section{Harmonic Space Formalism}\label{HarmonicAppendix}

Here we outline our methodology for the harmonic-space test presented in \secref{Clresults}.

The harmonic-space (often, simply `angular') power spectrum for the $i-j$ redshift bin pair is defined as
\begin{equation}
C_{\ell}^\hinospace(z_i,z_j)=4\upi\int\de\ln k\,\P_\zeta(k)\W_\ell^\hinospace(k;z_i)\W_\ell^\hinospace(k;z_j),
\label{eq:Cl_fullsky}
\end{equation}
with $\P_\zeta(k)$ the (dimensionless) power spectrum of primordial curvature perturbations. The redshift-integrated kernel for the $i$th redshift bin is given by
\begin{multline}
\W^\hinospace_\ell(k;z_i)=\T(k)\int\de\chi\,D(\chi)\Big[b_\hinospace(\chi)n^i(\chi)j_\ell(k\chi)
\Big],
\label{eq:weight_func}
\end{multline}
where: $\T$ is a transfer function, describing how gravity processes the growth of perturbations for different wave numbers; $\chi$ is the comoving distance to redshift $z$; $D$ is the growth factor, as in \autoref{eq:growth_factor}; and $j_\ell$ the $\ell$th-order spherical Bessel function.\footnote{Note that, for simplicity, we here disregard all other contributions to the observed angular clustering \citep[see e.g.][]{Bonvin:2011bg,Challinor:2011bk}.}

In the presence of PNG, when $b_\hinospace\to b_\hinospace+\Delta b_\hinospace=b_\hinospace+\widetilde{\Delta b}_\hinospace\,\fnl$, we can split the integral in \autoref{eq:Cl_fullsky} as follows,
\begin{multline}\label{eq:HarmonicCl}
    C_{\ell}^\hinospace(z_i,z_j)=
    C_{\ell}^{\hinospace,{\rm G}\times{\rm G}}(z_i,z_j)\\
    +2\fnl\,C_{\ell}^{\hinospace,{\rm G}\times{\rm NG}}(z_i,z_j)+\fnl^2\,C_{\ell}^{\hinospace,{\rm NG}\times{\rm NG}}(z_i,z_j),
\end{multline}
where `G' and `NG' respectively stand for `Gaussian' and `non-Gaussian'. The kernel for these terms reads
\begin{align}
    \W^{\hinospace,{\rm G}}_\ell(k;z_i)&=\T(k)\int\de\chi\,D(\chi)b_\hinospace(\chi)n^i(\chi)j_\ell(k\chi),\\
    \W^{\hinospace,{\rm NG}}_\ell(k;z_i)&=\T(k)\int\de\chi\,D(\chi)\widetilde{\Delta b}_\hinospace(k,\chi)n^i(\chi)j_\ell(k\chi).
\end{align}
As with the $P_\hinospace(k)$ investigation, we forecast this approach in the context of an SKA1-MID Band 1 survey (see information in \autoref{SKASurveyTable}). We generate 100 realizations of the \hi intensity map using the power spectrum from \autoref{eq:HarmonicCl} with $\fnl=0$. The theoretical inputs for $C^\hinospace_\ell$ are computed with a modified version of \texttt{CAMB\_sources},\footnote{\href{https://github.com/ZeFon/CAMB_sources_MT_ZF}{github.com/ZeFon/CAMB\_sources\_MT\_ZF}} first presented in \citet{Camera:2013kpa} and further upgraded in \citet{Fonseca:2015laa}. These fields are then smoothed with the beam, overlaid with Gaussian instrumental noise with variance $\sigma_\mathrm{noise}^2$ (\autoref{eq:Noise}), then lastly cut to a region with $20,000\,\text{deg}^2$.

The beam in harmonic-space can be modelled by
\begin{equation}
    B^2_\mathrm{b}(\ell) = \exp \left[\ell(\ell+1)\left(\theta_\mathrm{FWHM} / \sqrt{8 \ln 2}\right)^{2}\right]\,,
\end{equation}
and the uncertainty on the measured harmonic-space power spectrum is given by
\begin{equation}
    \delta C_\ell^\hinospace = \sqrt{\frac{2}{(2 \ell+1) \Delta \ell f_\mathrm{sky}}}\,(C_\ell^\hinospace + N_\ell)\,,
\end{equation}
where $N_\ell = \sigma_\mathrm{noise}^2\Omega_\text{pix}$.


\bsp	
\label{lastpage}
\end{document}